\documentclass[useAMS,usenatbib]{mn2e}
\usepackage{graphicx}
\usepackage{setspace}
\usepackage{natbib}
\usepackage{amssymb}
\usepackage{epsfig,color}
\usepackage{multicol}
\usepackage{epstopdf}
\DeclareGraphicsExtensions{.pdf,.eps,.png,.jpg,.mps}

\long\def\symbolfootnote[#1]#2{\begingroup%
\def\thefootnote{\fnsymbol{footnote}}\footnote[#1]{#2}\endgroup}

\renewcommand{\log}{\mathop{\mathrm{lg}}\nolimits}

\title[IMF from fragmentation and competitive accretion]
  {Stellar and substellar initial mass function: a model that implements gravoturbulent fragmentation and accretion}
\author[Veltchev, Klessen \& Clark]
  {Todor V.~Veltchev$^{1,2\,\star}$, Ralf S.~Klessen$^2$, and Paul C.~Clark$^2$\\
  $^1$University of Sofia, Faculty of Physics, 5 James Bourchier Blvd., 1164 Sofia, Bulgaria\\
  $^2$Institute of Theoretical Astrophysics, Albert-\"Uberle-Str. 2, 69120 Heidelberg, Germany}

\date{Submitted 2010 Xxxxx XX}

\pagerange{\pageref{firstpage}--\pageref{lastpage}} \pubyear{2010}
\def\LaTeX{L\kern-.36em\raise.3ex\hbox{a}\kern-.15em
    T\kern-.1667em\lower.7ex\hbox{E}\kern-.125emX}

\begin{document}

\label{firstpage}

\maketitle

\begin{abstract}
In this work, we derive the stellar initial mass function (IMF) from the superposition of mass distributions of dense cores, generated through gravoturbulent fragmentation of unstable clumps in molecular clouds (MCs) and growing through competitive accretion. MCs are formed by the turbulent cascade in the interstellar medium at scales $L$ from $100$ down to $\sim0.1~\rm pc$. Their internal turbulence is essentially supersonic and creates clumps with a lognormal distribution of densities $n$. Our model is based on the assumption of a  power-law relationship between clump mass and clump density: $n\propto m^{x}$, where $x$ is a scale-free parameter. Gravitationally unstable clumps are assumed to undergo isothermal fragmentation and produce protostellar cores with a lognormal mass distribution, centred around the clump Jeans mass. Masses of individual cores are then assumed to grow further through competitive accretion until the rest of the gas within the clump is being exhausted. The observed IMF is best reproduced for a choice of $x=0.25$, for a characteristic star formation timescale of $\sim5~\rm Myr$, and for a low star formation efficiency of  $\sim 10\,\%$. 
\end{abstract}


\section{Introduction}
The origin of the initial mass function (IMF) is a long-standing issue in modern astrophysics. It has numerous implications: from cosmology (e.g. cosmic reionization and formation of first galaxies), through studies of galactic structure and evolution, down to formation of planets and planetary systems. As such, an explanation of the IMF is one of the key goals of star formation (SF) research. It should reflect not only the variety of initial conditions in SF sites but implement complex and intertwined physical processes like gravitational collapse and fragmentation, turbulent motions, shock waves, accretion and protostellar outflows. 
\symbolfootnote[0]{$\star$~E-mail: eirene@phys.uni-sofia.bg}

The extensive photometric and spectroscopic studies of the Milky Way and some nearby galaxies over the last two decades led to reliable determination of the IMF in the range from $0.02-0.07~\rm M_{\odot}$ (brown dwarfs, BDs) up to $\sim 100~\rm{M}_\odot$. The high-mass part of the IMF ($\gtrsim1.0~\rm  M_\odot$) is nowadays firmly established to be a power-law function $dN/d\,{\log}\,M \propto M^\Gamma$. The universality of its `Salpeter slope' $\Gamma\simeq-1.35$ \citep{Sal55} is confirmed by infrared observations of the Arches cluster \citep{Kim_ea06, DKS07} although $\Gamma$ could vary ($\pm0.5$) in some regions of active star formation (see Elmegreen 2009 for a review). The low- and intermediate-mass IMF ($0.08 \lesssim M \lesssim 1~\rm M_\odot$) is a plateau of remarkable uniformity under various environmental conditions in the ISM \citep{EKW_08}. It could be represented with a power-law fit with much shallower slope \citep{Kroupa01} or with a lognormal function \citep{Ch03}. Significant uncertainties remain regarding the BD range of the IMF. For instance, \citet{TK07} demonstrate sensible discontinuity at $\sim0.08~\rm M_\odot$ in several young Galactic clusters.

Efforts dedicated both in analytical and semi-analytical studies and in numerical simulations led to obvious advance in reproducing the observational IMF in some Milky Way clusters but the picture is far from being complete (cf. Bonnell, Larson \& Zinnecker 2007 for review). One current theory describes the SF process as being controlled by supersonic turbulence \citep{PN02, MK04, BP_ea_06, HC08, HC09}. This process is often referred to as {\it gravoturbulent fragmentation}. The physics of turbulence is still quite poorly understood, because of the great mathematical complexity of the fluid equations (see e.g. Lesieur 1997) and the issue of how turbulence is driven remains under debate \citep{MK04}. Numerical simulations show that although supersonic turbulence can provide global support, it produces density enhancements in molecular clouds (MCs) that allow for local collapse - gravitationally bound clumps of a few hundred solar masses contract and fragment with formation of compact protostellar cores \citep{KB00, KB01, BBB03, C_ea05}.

If such forming clusters are gas-rich, {\it competitive accretion} is another crucial factor for the origin of the IMF -- it  determines the final mass of each protostellar core, depending on its initial mass and the properties of the surrounding turbulent flow \citep{Bon01b}. On the other hand, winds from the most massive stars play a crucial role for the energetic balance in the ISM of the cluster. Numerical simulations have also confirmed that radiation feedback from young protostars can also help shape the cluster IMF, by setting a lower limits to the Jeans mass in the ambient gas (e.g. Offner et al. 2009). Effects of accretion and energetic feedback from winds are implemented in a model of the cluster IMF of \citet{Dib_ea_10}.  

To sum up, current theories of the IMF describe the latter as the combined result of gravoturbulent MC fragmentation and competitive accretion, offering different scenarios, responsible for its substellar part \citep{BLZ07}. In this Paper we present a semi-analytical model that implements these two basic physical mechanisms and is in agreement with the generalised multipart power-law form of the observational IMF \citep{Kroupa01}. The `backbone' of our model is the mass distribution of gravitationally unstable, star-forming clumps which is linked to the probability density function (pdf) of the density field at different scales in MCs.  

\subsection{Influential (semi-)analytical works on the IMF}
The most successful (semi-)analytical IMF models of \citet{PN02} (hereafter: PN02) and of \citet{HC08} (hereafter: HC08) are based on the pdf of the density field, resulting from supersonic turbulence. The distribution of the density $n$ per unit volume is lognormal and the position of the maximum is a function of the standard deviation (stddev) $\sigma$: 
\begin{equation}
\label{eq_pdf}
p(z)\,d\,z=\frac{1}{\sqrt{2\pi \sigma^2}}\,\exp{\Big(-\frac{1}{2}\big( \frac{z -z_{\rm max}}{\sigma}\big)^2 \Big)}\,d\,z~,
\end{equation}
\begin{equation}
\label{max_pdf}
z_{\rm max}=-\frac{\sigma^2}{2}~,
\end{equation}
where $z\equiv{\rm ln}\,(n/n_0)$ and $n_0$ is mean density in the considered volume. The stddev depends on the Mach number $\cal M$ for supersonic flows \citep{PNJ97}:
\begin{equation}
\label{sigma_pdf}
\sigma^2={\rm ln}\,(1+b^2\,{\cal M}^2),~~~b\approx0.5
\end{equation}

In the semi-analytical model of PN02, the pdf of the density field is transformed into a distribution of Jeans masses. This approach serves for the probabilistic evaluation of the local collapse conditions for dense cores and hence to set up a low-mass cut-off of the IMF (see formula (24) in PN02). The cores themselves are identified as fragments of layers or filaments, formed by shocks in the supersonic turbulent flow. Their mass distribution is derived from the MHD shock jump conditions, whereas their number is controlled by a natural assumption for turbulent flows: self-similarity at different scales ($N_{\rm c}(L)\propto L^{-3}$). More recently, \citet{DKS07} extended the model of PN02, including dynamical evolution of the prestellar cores and their coalescence and collapse. The main criticisms to PN02 model concern the implementation of the geometry of the post-shock structures \citep{Elme07} and of turbulent and magnetic support in the collapse conditions (HC08). The mass of a formed dense core should also influence the suggested Jeans mass distribution. 

The extensive analytical model of HC08 is an attempt to refine the physical description of collapsing structures in the framework of \citet{PS74}. A density threshold is imposed for collapsing clumps while turbulent and magnetic support are implemented in the threshold conditions -- through dependence of the turbulent Jeans mass from the  mean turbulent velocity $\langle V_{\rm rms}^2\rangle^{1/2}$ and a modification of the Mach number in the MHD case, respectively. Scale dependence of the pdf width is introduced in the natural limits between injection and dissipation in a turbulent cascade process. Accretion onto the dense prestellar cores is assumed to happen only within the initial gravitationally unstable regions and not from external sources. That arises consistently from the picture of fragmentation in collapsing structures with power-law density profile $\rho\propto r^{-\alpha}$ ($\alpha < 2$). No further fragmentation during the core collapse is assumed.

Both mentioned models reproduce the high-mass, power-law part of the IMF with Salpeter slope $\Gamma\simeq-1.35$ from the implied turbulent theory. In PN02 this prediction is recovered from scale-dependence of $V_{\rm rms}$ and self-similarity at different scales, while in HC08 it is a substantial effect of inclusion of turbulent support. The IMF peak mass $M_p$ in PN02 model is a function of the average Bonnor-Ebert mass and scales inversely with the rms Mach number (or, the rms Alfv\'enic Mach number in the MHD treatment). Adopting typical values of the ISM parameters in star-forming cores\footnote{Mean column density $\sim10^{22}~\rm cm^{-2}$, $n\sim10^4~\rm cm^{-3}$, $T=10~\rm K$, magnetic field $B\sim10~\rm \mu G$; see formulae 9) and 10) in \citet{P_ea07}.}, one gets $M_p \simeq 0.16 - 0.25~\rm M_\odot$. In the approach of HC08, the peak of the IMF is a result of transition between the regime with significant turbulent support and the purely thermal behaviour. Its value is approximately constant ($M_p\sim0.1~\rm M_\odot$) in models with insignificant turbulent support (small local Mach numbers) and varies within a larger range ($ \sim 0.03-0.3~\rm M_\odot$) and toward lower masses in models with increasing supersonic turbulence.

\subsection{Main features of the presented model}
\label{our_model}
Our semi-analytical model attempts to combine some of the mentioned ideas about the origin of the IMF with several refined physical assumptions about SF process: 
\begin{itemize}
\item Starting point is a turbulent cascade process in the ISM, which for simplicity we assume to be isothermal at $T=10~\rm K$ at all scales. It leads to the formation of clumps at different scales $L$, beginning at an injection scale $L_{\rm inj}=100~\rm pc$ and proceeding downwards to $L_{0}\gtrsim0.1~\rm pc$. 
Mean density and velocity dispersion at each scale are estimated according to the Larson's (1981) relationships.
\item The clumps, generated through the supersonic turbulent flow at a given scale, have a lognormal density distribution that corresponds to the pdf of the density field. Assuming a power-law relationship between their mass and size, the clump mass distribution is obtained. 
\item Gravitationally unstable clumps are selected at scales $L_0\le L \le L_{\rm inj}$ and their composite clump mass function (CMF) is derived. Each of them fragments further and produces prestellar cores at the local Jeans scale. 
\item Competitive accretion onto the formed prestellar cores distributes the mass of the remaining gas from the fragmented clump. Clumps, which have transformed all their material to prestellar cores (by fragmentation and accretion), are replenished at constant rate.
\item Eventually, the IMF is derived from the composite CMF, assuming ongoing generation (replenishment) of clumps  and taking into account their fragmentation and the accretion on the formed prestellar cores.
\end{itemize}

In Section 2 we describe how the composite mass function of gravitationally unstable clumps is 
derived. Section 3 presents the derivation of the IMF from the CMF through fragmentation and accretion. We discuss the model predictions in Section 4 and summarise the main contributions of our model in Section 5.

\section{The clump mass function (CMF)}
\subsection{The concept of clump and its physical parameters}
\label{def_clumps}
Supersonic isothermal turbulence in MCs creates an intricate network of interacting shocks resulting in density fluctuations, described statistically by a lognormal pdf (Eq.~\ref{eq_pdf}). This is to be expected from the stochastical nature of turbulent flows and is confirmed from various numerical simulations \citep{PNJ97, Ost01, LKM03, Kri_ea_07, FKS08, Fed_ea_10}. To attribute a given density from the pdf to a spatial physical object (condensation) with particular volume and shape, one needs an identification scheme for structures in MCs. Common methods like CLUMPFIND \citep{WdB94} identify contiguous structures in datacubes with densities over a given threshold as distinct condensations, labeled generally ``clumps''. Usually, clumps are a broadly defined group of objects with sizes from several tenths of parsec\footnote{In order to be distinguished from dense prestellar cores with sizes $\lesssim0.1~pc$.} to $\lesssim 10$ pc and with masses in the range $10-10^4~\rm M_\odot$ \citep{Kra98, WBM00, Kauf_ea_10a}. They are, at least, confined by the external pressure of the ambient medium \citep{BM_92} and those which are virialised or gravitationally bound are the massive clumps where cluster formation takes place \citep{WBS95}.

Hereafter in this Paper, we use the term {\it clumps} for MC fragments of various shapes that have been formed by turbulent shocks at scales $0.1\lesssim L <100$~pc and have scale-dependent density distributions according to Eq.~\ref{eq_pdf}. They are potential SF sites where clusters of dense prestellar cores could form (Sect.~\ref{clump_fragm}). A statistical approach is adopted, assuming for simplicity that all clumps have cubic shape -- a clump size $l$ corresponds to clump volume of $l^3$. We suppose a power-law mass-size ($m-l$) relationship for clumps which also implies a mass-density ($n-m$) relationship:
\begin{equation}
\label{eq_n-m}
\frac{n}{n_0}=\Big(\frac{m}{m_0}\Big)^x=\Big(\frac{l}{l_0}\Big)^{C_x}~,~~~C_x=\frac{3x}{1-x},
\end{equation}
\begin{equation}
\label{eq_m-l}
\frac{m}{m_0}=\Big(\frac{l}{l_0}\Big)^{3/(1-x)},
\end{equation}
where $n_0(L)$, $m_0$ and $l_0$ are units of normalization. The existence of a mass-density (or, mass-size) relationship for molecular clouds and clumps can be expressed in terms of the combination of a velocity scaling law with different considerations of energy balance: equipartition of energies \citep{BP_VS_95}, virial equilibrium or relationships between the virial parameter or the Jeans number and the clump mass \citep{Dib_ea_07, Shet_ea_10}. Depending on the chosen approach, the result for the power exponent is in the range $-1.5\lesssim x\lesssim 0.4$ (its demonstration is beyond the scope of this paper). Relatively large variations of $x$ are obtained as well from present observational studies of cloud fragments in MCs ($-2\lesssim x \lesssim -0.5$, Kauffmann et al. 2010b) and numerical simulations ($-2\lesssim x \lesssim 0.$, Shetty et al. 2010). Therefore $x$ is taken to be a free parameter of our model. To avoid implementation of additional physics, scale independence of $x$ is also assumed although this might be a crude approximation.

\subsection{CMF at given scale}
\subsubsection{Clump density distribution}
As mentioned above (Sect.~\ref{our_model}), we use the proposed clump mass-density relation (Eq.~\ref{eq_n-m}) to derive the CMF from the clump density distribution. An appropriate statistical description of the latter is a lognormal pdf (Eq.~\ref{eq_pdf}). Its scale dependence is set by the scaling of the Mach number (cf. Eq.~\ref{sigma_pdf}), i.e. of the velocity dispersion $V_{\rm rms} (L)$, and by the choice of normalization unit $n_0 (L)$. Observational estimates of those quantities for the large range of considered scales could be provided from the so called ``Larson's laws'' \citep{L81}.
\begin{equation}
\label{Larson_v}
V_{\rm rms}=1.1\,\Big(\frac{L}{1\rm ~pc}\Big)^{a}~~\rm[km/s]~,~~~a\simeq0.4,
\end{equation}
\begin{equation}
\label{Larson_n}
n_0\equiv\langle n \rangle=0.34\,\Big(\frac{L}{\rm 1~pc}\Big)^{-1}~~\rm[10^4~cm^{-3}] 
\end{equation}
These relations do {\it not} imply anything about the specific parameters of the clumps themselves and their behaviour (e.g. the mass-density relationship) but are used here to give appropriate, statistically averaged values of $V_{\rm rms}$ and $n_0$ at each scale of the turbulent cascade. Choosing $n_0\equiv\langle n\rangle$ corresponds to setting a typical density in the considered volume, so that clumps could be thought as contiguous structures within isodensity contours in units $n_0$.

The other factor that determines the clump density distribution is the parameter $b$ (cf. Eq.~\ref{sigma_pdf}). Its value is subject to some uncertainty. Observations \citep{Brunt_10} and 3D numerical simulations \citep{PNJ97, Kri_ea_07} suggest $0.25 \le b\simeq 0.6$. However, \citet{FKS08} demonstrated that the type of turbulence forcing should be taken into account when $b$ is evaluated -- compressive forcing produces a three times larger stddev of the pdf than in the purely solenoidal case (see also Federrath et al. 2009). That corresponds to variations of $b$ in the range $0.33-1.0$ which we adopt further in this work. 

\subsubsection{Clump mass and size distributions}
\label{mass_and_size_dist}
After the scaling and the normalization of the lognormal clump density distribution have been specified, one is able to derive the mass and size distributions. Those are also lognormal, as Eq.~\ref{eq_n-m} imply, and we denote them $p_m({\rm ln}\,(m/m_0))$ and $p_l({\rm ln}\,(l/l_0))$, respectively. Their parameters are obtained straightforwardly:
\begin{equation}
\label{param_x}
z_m\equiv{\rm ln}\,(m/m_0)~:~~~z_{m,\,\rm max}=z_{\rm max}/x~,~~~~\sigma_m=\sigma/|x|
\end{equation}
\begin{equation}
\label{param_Cx}
z_l\equiv{\rm ln}\,(l/l_0)~:~~\,~~~~z_{l,\,\rm max}=z_{\rm max}/C_x~,~~~~\sigma_l=\sigma/|C_x|
\end{equation}

A lognormal pdf of the density field, generated by turbulence, has no natural limits; thus the derived clump mass and size distributions have not such. There is no natural choice for the normalization units $m_0$ and $l_0$ as well. We obtain the size and mass range and the normalization units by use of the requirements for volume and mass conservation:
\begin{equation}
\label{sum_l}
V_{\rm scale}=L^3=N \sum\limits_V l^3 (N_l/N)\simeq Nl_0^3\int e^{3z_l}\,p(z_l)\,d z_l
\end{equation}
\begin{equation}
\label{sum_m}
\langle\rho\rangle L^3=\mu n_0 L^3=M_{\rm scale}\simeq Nm_0^3\int e^{z_m}\,p(z_m)\,d z_m
\end{equation}
where $N_l$ is the number of clumps of size $l$, $N$ is the total number of clumps in the volume $V$ and the mean molecular weight is taken to be $\mu=2.4 m_p$. Consecutive numerical integration, performed symmetrically around $z_{l,\,\rm max}$, yields the size limits of clumps as lower and upper cutoffs when Eq. (\ref{sum_l}) and (\ref{sum_m}) are satisfied. The integrals on the right-hand sides have analytical solutions in the limits $(-\infty,+\infty)$. Using them, one obtains an approximate relation between the normalization units:
\begin{equation}
 \mu \frac{n_0 l_0^3}{m_0}\simeq\exp\Big(\sigma^2\cdotp\frac{1-x}{x}\Big)~,
\end{equation}
For the sake of numerical integration, it is appropriate to choose $l_0$ to be small and proportional to scale size $L$ (of order of few percent of it):
\[ l_0=k_l\,L \]

\begin{figure*} 
\includegraphics[width=1.\textwidth]{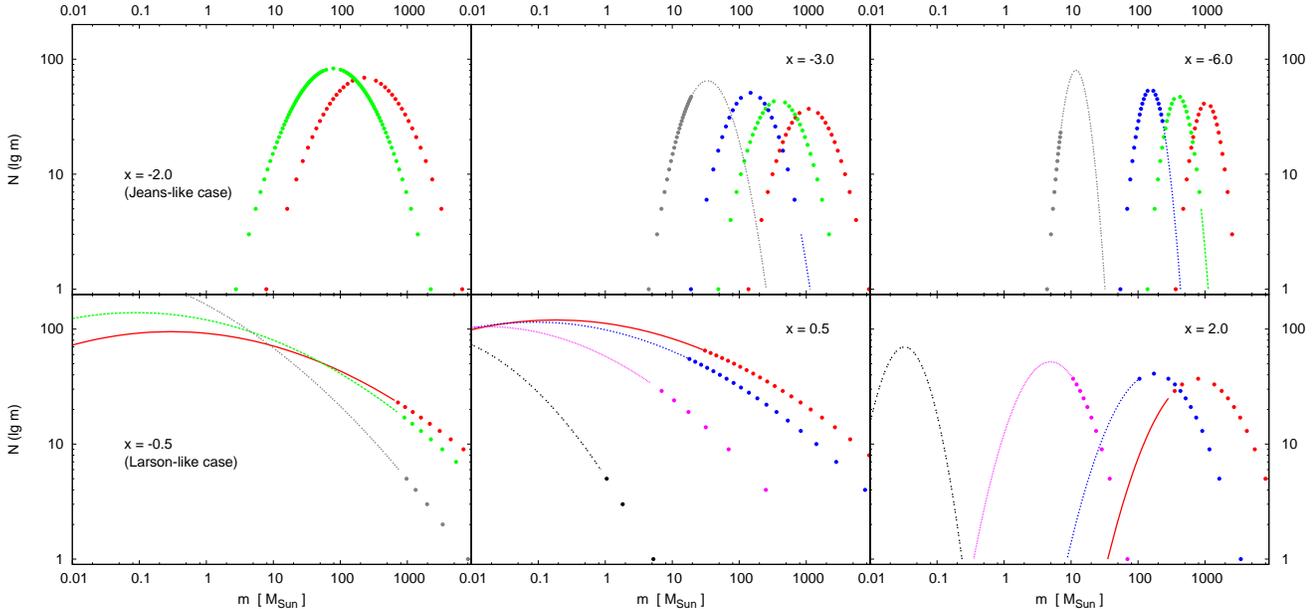}
\caption{Clump mass distributions for different values of $x$ and at different scales: $100~\rm pc$ (red), $70~\rm pc$ (green), $50~\rm pc$ (blue),  $30~\rm pc$ (grey), $10~\rm pc$ (violet) and $1~\rm pc$ (black). The forcing parameter is fixed: $b=0.5$. The portions that correspond to gravitationally unstable clumps are shown with bullets.}
\label{m_pdfs} 
\end{figure*}

\subsubsection{Mass distribution of unstable clumps}
\label{mass_distr_jeans}
Clumps for which the mass exceeds the local Jeans mass $m_{\rm J}(n)$ are, by definition, gravitationally unstable. Their fraction or formation at given scale $L$ depends on some {\it critical mass} $m_{\rm cr}(L)$ that separates the regimes $m/m_{\rm J}(n)=m/m_{\rm J}(m)\ge1$ and $m/m_{\rm J}(m)<1$.
 
Introducing the chosen normalization unit $n_0\equiv\langle n \rangle$ into the Jeans mass expression $m_{\rm J}\simeq1.5\,(T/10~K)\,(n/10^4~\rm cm^{-3})$ and using the mass-density relationship (Eq.~\ref{eq_n-m}), we obtain the criterion for gravitational instability of a clump:
\begin{equation}
\label{inst_cl}
\frac{m}{m_{\rm J}}=K(x,L)\,\Big(\frac{L}{\rm 1~pc}\Big)^{-\frac{1+2x}{2}}\Big(\frac{m}{\rm M_\odot}\Big)^{\frac{x+2}{2}}\ge1 
\end{equation}
which yields a critical mass:
\begin{equation}
\label{lim_inst_cl}
m_{\rm cr}=K(x,L)^{-\frac{2}{x+2}}\,\Big(\frac{L}{1~\rm pc}\Big)^{\frac{1+2x}{x+2}}~~\rm [M_\odot]
\end{equation}
The factor $K(x,L)$ varies within an order of magnitude for $x>0$ and about 2 orders of magnitude for $x\le-3$. A brief analysis of Eq.~\ref{inst_cl} and \ref{lim_inst_cl} displays two special physical cases. 
\begin{itemize}
 \item $x=-0.5$: That is the {\it Larson-like case} in which $n\propto l^{-1}$ (cf. Eq.~\ref{eq_n-m} and \ref{Larson_n}). The critical mass is huge ($\sim 700-1500~M_\odot$) and depends weakly on the scale:
\begin{equation}
\label{mj_larson}
m_{\rm cr}=K(x,L)^{-\frac{4}{3}}
\end{equation}
 \item $x=-2$: All formed clumps are unstable by definition ($m\propto n^{-1/2}$) and an appropriate name is {\it Jeans-like case}. It is obviously unrealistic, although the mass range of unstable clumps is significantly larger than in the Larson-like case. The criterion for gravitationally instability (Eq.~\ref{inst_cl}) is fulfilled only for scales where $K(x,L)>(L/{\rm1~pc})^{-3/2}$, or $L\gtrsim70~\rm pc$. If realised (partially) in nature, the Jeans-like case would correspond to starburst regions at larger galactic scales.
\end{itemize}

The clump mass distributions at different scales, produced in the above cases, are shown in Figure~\ref{m_pdfs} (left panels). Two other cases worth mentioning are:
\begin{itemize}
 \item[$\ast$] $x=0$: Homogeneous medium with no density fluctuations ({\it case of decayed turbulence}).
 \item[$\ast$] $x\simeq1$: Degeneration in the clump size - clump density relation (Eq.~\ref{eq_n-m}), i.e. clumps of approximately constant size have densities in an extremely large range.
\end{itemize}
The clump mass distributions for some other values of $x$ are plotted in Fig.~\ref{m_pdfs} (centre and right panels). Apparently, turbulent cascade in clouds of sizes $1-100~\rm pc$ gives birth to gravitationally unstable clumps in a large range of masses $1-10^5~\rm M_\odot$. A lower mass limit of $<1~\rm M_\odot$ is achieved for positive $x\lesssim0.5$. Negative values of $x$ below the Jeans-like value ($<-2.$) condition that almost all clumps that would form at large scales ($L\gtrsim80~\rm pc$) are unstable. That leads to a top-heavy composite CMF we derive in the next Section.

\begin{figure} 
\begin{center}
\includegraphics[width=84mm]{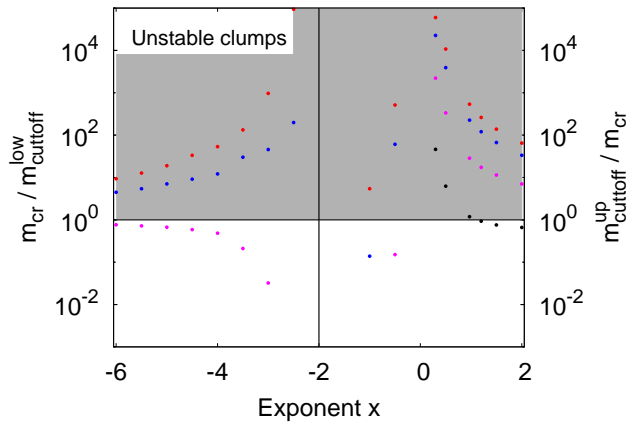}
\caption{Mass ranges of gravitationally unstable clumps in the regimes separated by the Jeans-like case: $x<-2$ (left) and $x>-2$ (right). Unstable clumps form if $m_{\rm cr}>m_{\rm cuttoff}^{\rm low}$ for $x<-2$ and $m_{\rm cuttoff}^{\rm up}>m_{\rm cr}$ for $x>-2.$. The scales are designated like in Fig.~\ref{m_pdfs}.}
\label{mmj_x}
\end{center}
\end{figure}

The abundance of gravitationally unstable clumps depends not only on the scale $L$ (Eq.~\ref{lim_inst_cl}), but also {\it on the mass distribution parameters:} the position of the maximum and the width (Eq.~\ref{max_pdf}, \ref{sigma_pdf} and \ref{param_x}). The latter dependence becomes stronger as $|x|$ grows -- $z_{max,\,m}$ at a given scale shifts to lower masses while $\sigma_m$ decreases (see Fig.~\ref{m_pdfs}). In such cases and especially at small scales, shape and position of the mass distribution are the crucial factor that determines whether the formation of unstable clumps would occur or not. In other words, the formation and the fraction of unstable clumps at given scale $L$ are a product of interplay between the values of $x$, $m_{\rm cr}$ and $\sigma_m$. This is illustrated in Fig.~\ref{mmj_x}. Unstable clumps form in the mass range:
\begin{equation}
\label{uclumps_xm}
m_{\rm cr}(L)\ge m >m_{\rm cutoff}^{\rm low}(x,L),~~~x<-2,
\end{equation}
\begin{equation}
\label{uclumps_xp}
m_{\rm cr}(L)\le m <m_{\rm cutoff}^{\rm up}(x,L),~~~x>-2,  
\end{equation}
where mass cutoffs $(m_{\rm cutoff}^{\rm low},m_{\rm cutoff}^{\rm up})$ are determined from simultaneous numerical integration of Eq.~\ref{sum_l} and~\ref{sum_m} (Sect.~\ref{mass_and_size_dist}). If $m_{\rm cr}(L)<m_{\rm cutoff}^{\rm low}(x,L)$ (or $m_{\rm cr}(L)>m_{\rm cutoff}^{\rm up}(x,L)$) for given pair $(x,L)$, no unstable clumps are formed. As seen in Fig.~\ref{mmj_x}, small positive values of $x$ yield lower scale limits of formation of unstable clumps $L_{0}(x)$: $L_{0}\simeq1~\rm pc$ for $x\simeq0.5$ and a value of $L_{0}\sim 0.1~\rm pc$, typical for transition from clumps to dense cores, is achieved only for $0<x\lesssim0.25$. On the other hand, negative values of $x$ lead to formation of unstable clumps only at scales $>10~\rm pc$. Noteably, turbulent flows remain supersonic at all scales, for all choices of $x$. 

\subsection{Composite CMF}
\begin{figure} 
\includegraphics[width=72mm]{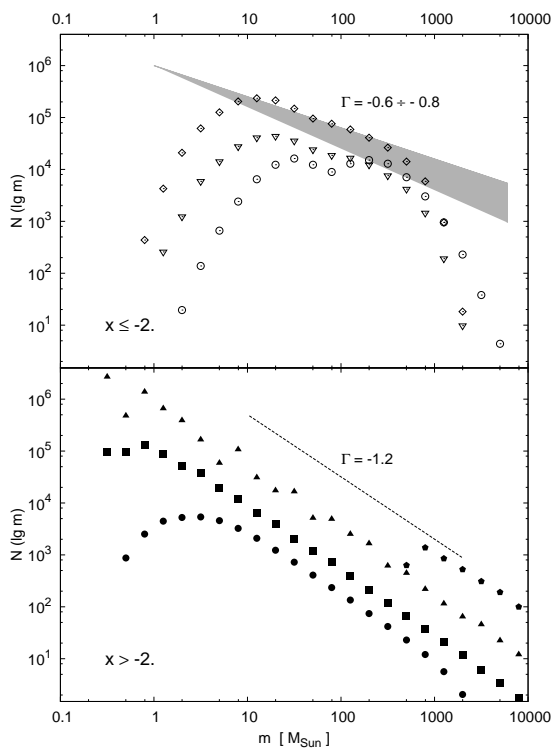}
\caption{Composite CMFs, derived for $x\le-2.$ (top, open symbols) and $x>-2.$ (bottom, filled symbols) values of $x$, like in Fig.~\ref{m_pdfs}: $-2.0$ (Jeans-like case; circles), $-3.0$ (triangles), $-6.0$ (diamonds); $-0.5$ (Larson-like case; pentagons), $0.25$ (squares), $0.5$ (squares), $2.0$ (circles). The CMFs in the bottom panel are artificially shifted vertically for clarity. Typical range of slopes (grey area) for observational CMFs \citep{Kra98} and the best-fit slope for $x\ge-0.5$ (dashed line) are plotted. } 
\label{cmfs} 
\end{figure}

The obtained mass distributions of unstable clumps have to be summed up over the range of scales $L_0(x)<L<L_{\rm inj}$ to derive a composite CMF, representative for star formation at galactic scales. In fact, aggregates of MCs of different sizes are embedded in the general ISM. If all of them are generated by supersonic turbulence through a cascade process, so a self-similar distribution in a given volume unit can be assumed. Hence the total number of clumps $N\propto L^{-3}$ (see PN02) and the contribution of clumps, generated at scale $L$, to a logarithmic mass bin $d(lg\,m)$, scales as
\begin{equation}
 N (L,~d(lg\,m))=\Big(\frac{L_{\rm inj}}{L}\Big)^3\,N (L_{\rm inj},~d(lg\,m))~~,
\end{equation}
where $L_{\rm inj}$ is the chosen injection scale of the turbulent cascade. Then, the total number of clumps, contributing to a mass bin, is:
\[ N_{\rm tot}\,d(lg\,m)=\int\limits_{L_{0}(x)}^{L_{\rm inj}}\,N (L,~d(lg\,m))\,dL  \]
The derived composite CMFs for a set of exponents $x$ are plotted in Fig.~\ref{cmfs}. Two different types of behaviour are apparent. For negative values of $x$, implying a steeper density-size relation than in ``Larson's second law'' ($n\propto l^a,~a<-1$), the CMFs are top-heavy, asymmetric, with steep edges and shallow power-law parts. The slopes are in the range $-0.6 \ge \Gamma \ge -0.8 $ , in agreement with observations of MC clumps, associated with active SF regions (e.g. Kramer et al. 1998), but the turn-over mass $M_{\rm to,\,CMF}$ is significantly larger, by an order of magnitude. The cases $x > -2$ yield smooth, power-law CMFs spanning a wide range of masses and with slope $\Gamma \approx -1.2$, close to the Salpeter value. The turn-over mass is $\lesssim1~M_\odot$ for small positive values of $x$ and depends on the low-mass limit of the CMFs. The latter is determined by the lowest scale of unstable clump formation $L_{0}(x)$. Therefore, in our approach, $M_{\rm to,\,CMF}$ is a scale-averaged, model-dependent (on the value of $x$) and {\it not} an universal phenomenon. 

\section{From the CMF to the IMF}
The derivation of the IMF from the CMF reflects the processes of fragmentation and further accretion on the generated protostellar cores. Hereafter, we will use the term {\it prestellar cores} (or, simply, cores) to denote those objects that form via the fragmentation of unstable clumps, and denote their density and mass with $n^\prime$ and $m^\prime$, respectively.

\subsection{Fragmentation of the unstable clumps}
\label{clump_fragm}
The most straightforward approach to fragmentation is to assume that each unstable clump of mass $m$ fragments into $(m/m_{\rm J})$ protostellar cores. That procedure, however, is not well justified physically due to, at least, two reasons: (1) it implies unrealistically high SF efficiency; (2) it neglects the internal structure and dynamics of a collapsing clump. \citet{Good08} adopt a purely probabilistic approach to describe fragmentation of low mass clumps ($0.1-100~M_\odot$) into 2 or 3 fragments, assuming constant ratios of binaries to triples and binaries to single stars, constant SF efficiency and an uniform probability distribution. We consider clumps in a significantly larger mass range and, in most cases, with $(m/m_{\rm J})\gg1$. Therefore we take a different approach on fragmentation.

In view of the fractal nature of turbulence, it is reasonable to describe clump fragmentation (also) through a lognormal core mass distribution. This should be a distribution of {\it Jeans masses}, i.e. local density contrasts $n^\prime/n(m_{\rm J})$ within the clump generate colapsing cores with corresponding Jeans masses $m^\prime=m_{\rm J}^\prime \propto (n^\prime)^{-1/2}$. The peak of the distribution is assumed to be at the clump Jeans mass $m_{\rm J}(n)$:
\[ m=\int\limits_{m_{\rm J}-\Delta m(\sigma)}^{m_{\rm J}+\Delta m(\sigma)}\!\!\!\!\!\!m_{\rm J}^\prime\,N(m_{\rm J}^\prime)\,dm_{\rm J}^\prime~;~~
N(m_{\rm J}^\prime)=\frac{p(m_{\rm J}^\prime)}{p(m_{\rm J}+\Delta m(\sigma))}  \]
where $\Delta m(\sigma)$ gives the numerical integration cutoffs. The width of the cores' mass distribution $\sigma_{m^\prime}=\sigma_{n^\prime}/2$ ($x=-2$) is calculated from Eq.~\ref{sigma_pdf}. In that equation, the global-scale Mach number ${\cal M}(L)$ has to be replaced by the local Mach number ${\cal M}_l$, reflecting the velocities $v_{\rm rms}$ at scales of the clump size $l$ (see comments on that in \citet{BB06}, Sect. 2). Clump sizes in our model span the range $0.01\lesssim l \lesssim 10~\rm pc$ which includes a variety of different physical regimes. Turbulence dominates the observed linewidth down to sizes of $l \sim0.1-0.2~\rm pc$ \citep{BGodd98}. Its scaling in that regime obeys relation: $v_{\rm rms} \sim l^{a}$, with a slope $a\simeq0.41-0.43$ \citep{P_ea06, P_ea09}, close to Larson's value (Eq.~\ref{Larson_v}), or steeper: $0.48 \le a \le 0.75$ \citep{HWB06}. At sizes $0.1-0.2~\rm pc$ the thermal component of $v_{\rm rms}$ becomes comparable to the turbulent one and the slope $a$ decreases to $0.1-0.15$. Eventually, at scales $\sim0.04~\rm pc$, ``coherent cores'' with $v_{\rm rms}\approx \rm const$ are observed \citep{AGood98}. An appropriate parametrization of the velocity dispersion in the fragmenting clumps, describing a smooth transition between turbulent and thermal regimes, is:
\begin{equation}
 v_{\rm rms}^2(l)=v_0^2\,\big(C+(l/l_0)^{2a}\big)~~,
\end{equation}
where we take $v_0=0.8$ and $a=0.48$, obtained for MC with low SF efficiency \citep{HWB06}, and fiducial size $l_0=0.8~\rm pc$. $C$ is a small constant for which we choose a value of $0.1$.

Similar to \citet{Good08}, we define {\it prestellar core formation efficiency} $\epsilon$ (PCFE) as the percentage of clump mass $m$ that remains bound in the cores, after the initial fragmentation. According to observational estimates, the total gas mass in young stellar clusters is typically many times the mass in stars \citep{Lada91}. Therefore it is reasonable to choose small values of $\epsilon$ {(see also Krumholz \& Tan 2007, ApJ)}. To simplify the calculations, we adopt constant values of this parameter, although one can expect that it is scale-dependent \citep{VS03}. The rest of the initial clump mass $m_{\rm gas}(t_0)=(1-\epsilon)m$ is distributed further among the formed cores through competitive accretion. 

\subsection{Description of accretion on the prestellar cores}
\label{acc_descr}
Accretion in protostellar clusters proceeds in two main phases: gas-dominated potentials and stellar-dominated potentials \citep{Bon01b}. In our approach, small values of $\epsilon$ imply that the gas dominates the potential within the fragmented clump at some initial moment $t_0$. Description of accretion on cores under such conditions is difficult to implement in a (semi-)analytical approach. The general formula for the accretion rate on a protostar of mass $M$ is:
\begin{equation}
\label{acc_gas}
\dot{M}\approx \pi \rho v_{\rm rel}R_{\rm acc}^2~~,
\end{equation}
where $\rho$ is the gas density, $v_{\rm rel}$ is the relative gas-protostar velocity and $R_{\rm acc}$ is the accretion radius. If gas dominates the potential and the collapse is isothermal, a good estimation of $R_{\rm acc}$ is the tidal-lobe radius $R_{\rm tidal}$ \citep{Bon01a}. However, the evaluation of $R_{\rm tidal}=f(r)$ for each core requires knowledge of its current position $r$ within the clump, i.e. further assumptions about cores' spatial distribution and its evolution in time. Also, tidal accretion tends to dominate in the fragmentation phase (as described in the previous Subsection), not in the N-body dynamics that takes place afterwards \citep{BCB08}. It is therefore more appropriate to use the Bondi-Hoyle accretion description \citep{BH41}, devised for stellar-dominated potential:

\begin{equation}
\label{rad_BH}
R_{\rm acc}\approx R_{\rm BH}\approx \frac{2GM}{v_{\rm rel}^2+c_{\rm s}^2}
\end{equation}
\begin{equation}
\label{acc_BH}
\dot{M}\approx \dot{M}_{\rm BH}\approx 4\pi \rho \frac{(GM)^2}{(v_{\rm rel}^2+c_{\rm s}^2)^{3/2}}~~,
\end{equation}
where the gas density in protostar's vicinity is assumed to be uniform. The latter assumption is not far from reality also in a moving gas medium, if the gas distribution retains its general form with time \citep{Bon97}. The density profile $\rho (r,t)$ within a collapsing clump approaches quickly the isothermal form ($\rho\propto r^{-2}$), with a small nucleus of nearly uniform density \citep{L69}. One can adopt an approximation of the spherically averaged gas density profile in fragmented and collapsing clumps:
\begin{equation} 
\label{rho_t}
\rho(r,t)=\left\{ \begin{array}{ll}\rho_{\rm c}(t) & ~~~~~r\le r_{\rm c}(t)\\
	 \rho_{\rm c}(t)\,\big(r_c(t)/r\big)^2 & ~~~~~r>r_{\rm c}(t)
\end{array} \right.
\end{equation}
where $r_{\rm c}(t)$ is the radius of the clump nucleus with uniform density $\rho_{\rm c}(t)$ at a fixed moment of time. The time evolution of the latter quantity is approximately linear \citep{L69}:
\begin{equation}
\label{rhoc_t}
\rho_{\rm c}(t)=\rho_{\rm c}(t_0)+A\,t~~~,
\end{equation}
whereas $r_{\rm c}(t)$ could be obtained from the equation of gas mass $m_{\rm gas}(t)=\int 4\pi\rho(r,t)r^2\,dr$ and hence depends on the accretion rates within the clump. Then the mean density within a clump of size $l$ and at a fixed moment $t$ is:
\begin{equation}
\label{rho_mean}
\bar{\rho}(t)=\frac{1}{l}\int\limits_0^l \rho(r,t)\,dr=2\rho_{\rm c}(t)\,\frac{r_{\rm c}(t)}{l}\,\Big(1-\frac{r_{\rm c}(t)}{l}\Big)
\end{equation}
This estimate is physically more correct than to assume decreasing (due to accretion) uniform mean density. The relative gas-to-protostar velocity could be derived from the clump Mach number ${\cal M}_l$ (see previous Subsection). Eventually, accretion on a protostellar core of mass $m^\prime$ is described in our approach through the Bondi-Hoyle formula:
\begin{equation}
\label{acc_rate}
\dot{m^\prime}(t)=4\pi \bar{\rho}(t) \frac{(Gm^\prime)^2}{(v_{\rm rel}^2({\cal M}_l)+c_{\rm s}^2)^{3/2}}
\end{equation}
and the total mass of the gas within the clump is being exhausted at a rate $\dot{m}_{\rm gas}=\int \dot{m^\prime}\,N(m^\prime)\,dm^\prime$.

\subsection{Derivation of the composite IMF}

\subsubsection{Parameters and timescales}
\label{comp_IMF_param}
The derived IMF depends on three basic parameters. Two of them determine the CMF: the exponent $x$ in the clump mass-density relationship and the turbulent forcing parameter $b$ (Eq.~\ref{sigma_pdf}); while the third, the PCFE $\epsilon$, acounts for fragmentation.

\begin{figure} 
\includegraphics[width=84mm]{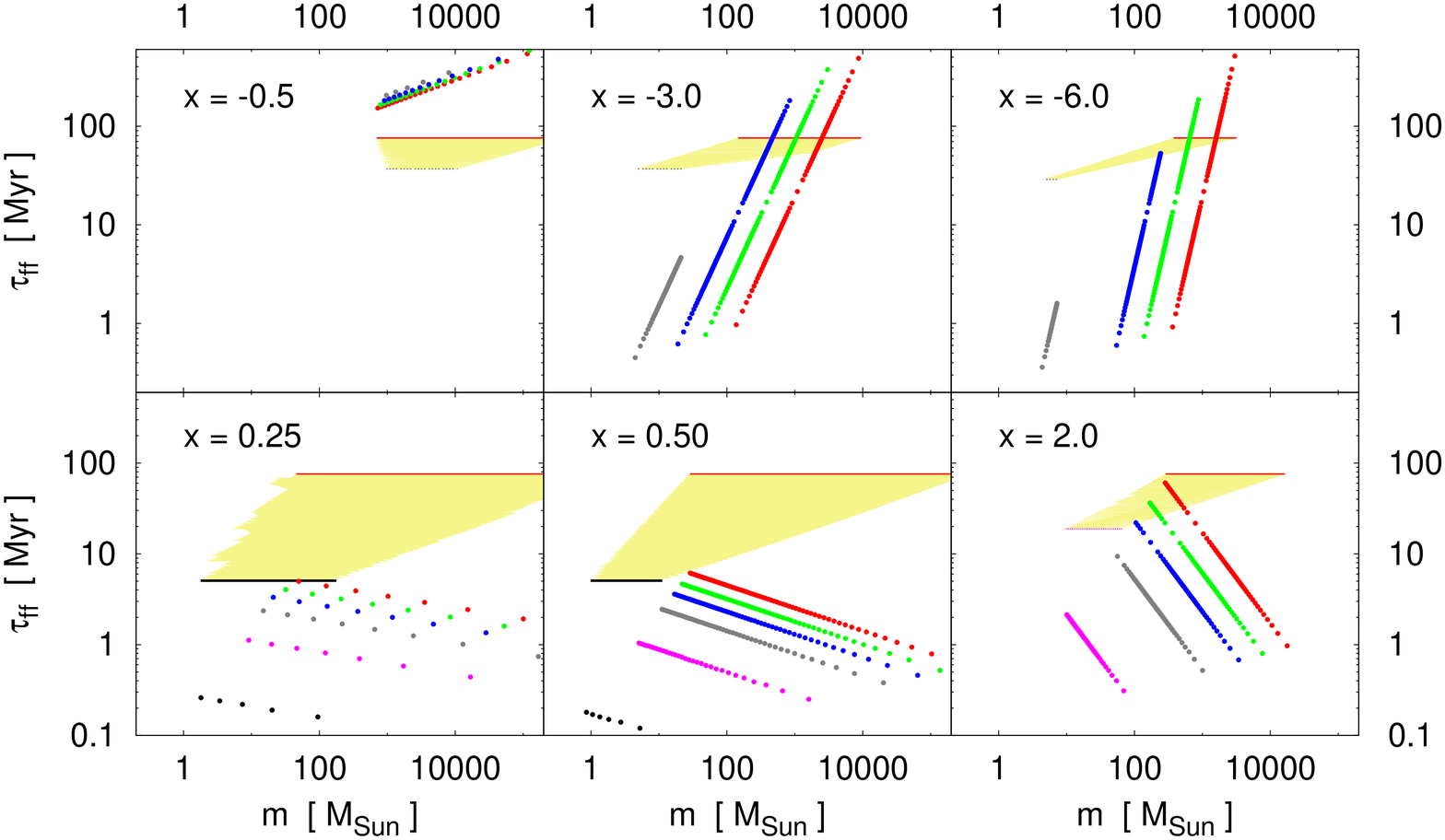}
\caption{Free-fall times of unstable clumps produced at scales 100, 70, 50, 30, 10 and 1 $\rm pc$ (designated as in Figs. 1 and 2) and for different values of $x$. Note the weak scale dependence for $|x| \lesssim 0.5$. The range of turbulent crossing times $\tau_{\rm cr}(L)$ is drawn for comparison (yellow area).} 
\label{m_tau} 
\end{figure}

The free-fall time $\tau_{\rm ff}$ of a unstable clump is a natural measure for description of its evolution (fragmentation and internal accretion). Since
\[ \tau_{\rm ff} = \Big( \frac{3\pi}{32 G\rho}\Big)^{1/2} \propto n^{-1/2} \propto m^{-x/2}~~, \]
the range of clump evolution timescales is determined by the clump mass distribution and, hence, from parameters $x$ (Eq.~\ref{param_x}) and $b$ (Eq.~\ref{sigma_pdf}). The maximal value $\tau_{\rm ff,\,max}$ is achieved at the largest (injection) scale $L_{\rm inj}=100~\rm pc$ and it could be taken as a characteristic evolution time of protostellar cores generated through the turbulent cascade. The dependencies of $\tau_{\rm ff}$ on clump mass $m$, turbulent scale $L$ and on $x$ are illustrated in Fig.~\ref{m_tau}. Larger scales produce wider spans of free-fall times and this correlation is very sensitive to the chosen value of $x$. For $|x|\gtrsim 1$, unstable clumps that are produced at scales $L\gtrsim50~\rm pc$ would evolve at free-fall times varying within two orders of magnitude! Moreover, $\tau_{\rm ff}(m,L)$ are comparable to or even greater (in the Larson-like case) than the turbulent crossing time at the given scale $\tau_{\rm cr}(L)\propto L^{0.6}$. That makes models with such values of $x$ implausible for description of the SF process. On the other hand, the cases $0.2 \lesssim x \lesssim 0.5$ produce a range of clump free-fall times from few tenths to several Myr, which is consistent with the rapid SF model, in its observational or theoretical timescales (Ballesteros-Paredes, Hartmann \& Vazquez-Semadeni 1999; Pringle, Allen \& Lubow 2001). 

The observational IMF at galactic scales is representative of stars born at different scales and hence within a large range of clump free-fall times $\tau_{\rm ff}\,(n,L)$. In some of the protostellar clusters the gas is exhausted and hence the accretion is halted. Others are still embedded in their natal cloud which sets certain observational limits to their detection and/or estimation of their masses. In this study, we abstain from implementing such observational aspects in the model and take into account only fragmented clumps (protoclusters) in our derivation of the IMF with initial mass $\epsilon m$ where the accretion has halted due to gas exhaustion at a moment of time $t_{\rm acc}$, less or comparable to $\tau_{\rm ff}$. The SF efficiency (SFE) at fixed time $t$ is defined as the ratio of the total (initial + accreted) mass of the prestellar cores $M_{\ast} (t)$ to the total mass $M_{\rm tot}$ involved in the whole turbulent cascade. The SFEs, obtained for a set of values of $x$ and $\epsilon$ and at $t=\tau_{\rm ff,\,max}$, are given in Table~\ref{tau_sfe}.

\begin{table}
\begin{center}
\caption{Maximal clump free-fall times and SFEs.}
\label{tau_sfe} 
\begin{tabular}{@{}ccrccc}
\hline 
\hline 
$x$ & $b$ & $\tau_{\rm ff,\,max}$  & \multicolumn{3}{c} {SFE at $t=\tau_{\rm ff,\,max}$ } \\ 
~ & ~ & [Myr] & $\epsilon=0.10$ & $\epsilon=0.40$ & $\epsilon=0.70$\\ 
\hline 
0.25 & 0.33 & ~5.4 & 0.12 & 0.51 & 0.89 \\ 
0.50 & 0.33 & ~7.5 & 0.56 & 0.75 & 0.89 \\ 
0.97 & 0.33 & 13.4 & 0.20 & 0.36 & 0.60\vspace{4pt}\\ 
0.25 & 1.00 & ~5.0 & 0.12 & 0.66 & 0.90 \\ 
0.50 & 1.00 & ~5.8 & 0.30 & 0.55 & 0.79 \\ 
0.97 & 1.00 & 15.0 & 0.20 & 0.35 & 0.58\vspace{4pt}\\ 
\hline 
\hline 
\end{tabular} 
\end{center}
\smallskip 
\end{table}

\subsubsection{Treatment of the timescale problem}
\label{treat_timescale}
To account for the various accretion timescales $t_{\rm acc}$ within the clumps in derivation of the IMF, we follow an approach discussed by \citet{CKB07}. Without any consideration of fragmentation and accretion, those authors demonstrate that the IMF would differ substantially in form from the progenitor CMF (existing over timescale $\tau_{\rm CMF}$), if the evolution timescale for the clump $t_{\rm evol}$ depends on its mass $m$: $f_{\rm IMF}\approx f_{\rm CMF}\times\tau_{\rm CMF}/t_{\rm evol} (m)$. In our model, stellar progenitors are the cores within a fragmented clump, with protostellar core mass function (PCMF) $f_{\rm PCMF}(\epsilon m)$ and evolution timescale $t_{\rm acc}(\epsilon m)$. Permitting the clump population to be constantly replenished, we have for the local IMF (LIMF) at $t>t_{\rm acc}(\epsilon m)$, produced from the PCMF through competitive accretion:
\begin{equation}
 f_{\rm LIMF}(t, L)=\frac{t}{t_{\rm acc}(\epsilon m)}\,f_{\rm PCMF}(\epsilon m, L)
\end{equation}

Eventually, the composite IMF at a fixed moment of time $t$ is derived through integration over the range of scales where unstable clumps are formed: 
\begin{equation}
 f_{\rm IMF}(t)=\int\limits_{L_{0}(x)}^{L_{\rm inj}}\,f_{\rm LIMF}(t, L)\,dL
\end{equation}

\section{Results and discussion}
\begin{figure} 
\includegraphics[width=84mm]{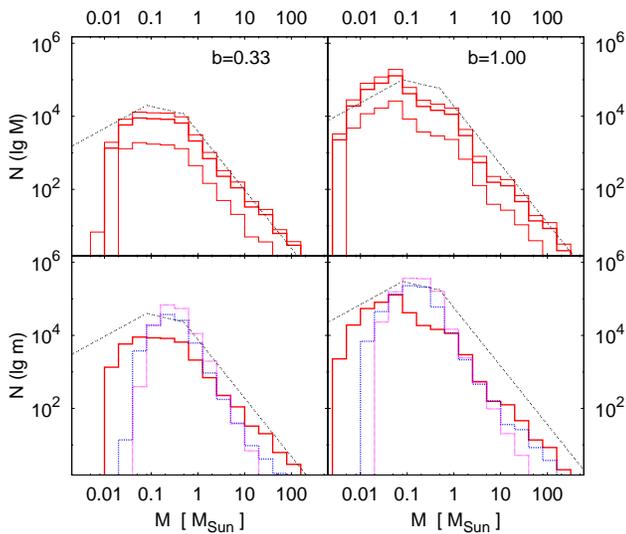}
\caption{Initial mass function for $x=0.25$. Results for purely solenoidal forcing (left panels) and for purely compressive forcing (right panels) are plotted: a) (top) fixed, low PCFE $\epsilon=0.1$ and time evolution $0.2, 1.0, 1.5\,\tau_{\rm ff,\,max}$ ; b) varying PCFE (bottom): $\epsilon=0.1$ (red), $\epsilon=0.4$ (blue), $\epsilon=0.7$ (violet), at $t=\tau_{\rm ff,\,max}$. Multipart power-law IMF according to \citet{Kroupa01} is plotted for comparison.} 
\label{imf_025} 
\end{figure}

\begin{figure} 
\includegraphics[width=84mm]{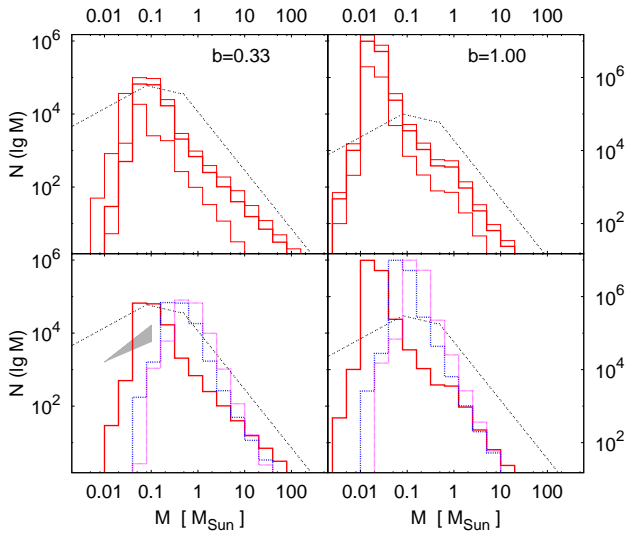}
\caption{The same like Fig.~\ref{imf_025}, but for $x=0.50$. The range of slopes of the BD-like IMF as obtained by \citet{TK07} is drawn as grey area.} 
\label{imf_050}
\end{figure}

\begin{figure} 
\includegraphics[width=82mm]{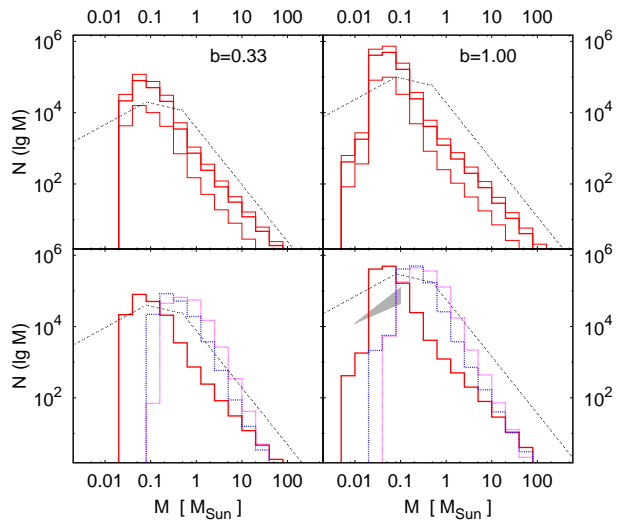}
\caption{The same like Fig.~\ref{imf_050}, but for $x=0.97$.} 
\label{imf_097} 
\end{figure}

Plausible evolution timescales of the protostellar clumps ($\sim\tau_{\rm ff}(x)$) are obtained for the slopes of the clump mass-density relation (Eq.~\ref{eq_n-m}) in the range $0<x\lesssim1$ (cf. Table~\ref{tau_sfe}). Therefore we focus on the results for the IMF in the cases $x=0.25$, $x=0.5$ and $x=0.97\approx1$), plotted in Figs.~\ref{imf_025}-\ref{imf_097}. Several apparent features should be pointed out:
\begin{itemize}
 \item A Salpeter slope of the high-mass IMF is generally reproduced for small PCFE $\epsilon$ for all choices of the exponent $x$. This result is not trivial since it is {\it not} a direct effect of transforming the CMF with a similar slope (cf. Fig.~\ref{cmfs}), but includes the effects of fragmentation and accretion. On the other hand, steeper slopes are obtained as $\epsilon$ is increasing. That is understandable, because decreasing of the initial gas mass $m_{\rm gas}$ in the clumps minimises the role of accretion on the protostellar cores in reshaping their mass distribution. 
 \item Intermediate-mass IMF with shallow (negative) slope and a mass range within an order of magnitude ($0.06 \lesssim M \lesssim 0.6~M_\odot$) is derived mainly for solenoidal turbulent forcing. Compressive forcing ($b=1.0$) tends to produce a narrow IMF, with a peak shifted toward the BD mass range. The apparent exception is the case $x=0.25$: both extreme regimes of forcing yield an IMF in a good agreement with observations.
 \item A discontinuity of the IMF is evident in the BD mass range for $x=0.50,~0.97$ and higher PCFE (Fig.~\ref{imf_050} and \ref{imf_097}). It is about one order of magnitude or less higher than what is found by  \citet{TK07} in the young clusters Trapezium and IC 348.
\end{itemize}
The model with $x=0.25$ and low PCFE exhibits the best consistency with the observational IMF. According to Eq. (\ref{eq_n-m}), this case corresponds to an interesting clump geometry -- its density is exactly proportional to the linear size: $n \propto l$. This result is in apparent discrepancy with  ``Larson's second law'' (Eq.~\ref{Larson_n}). It should be considered cautiously but not automatically rejected. The relation $n \propto l^{-1}$ (or, equivalently, $m \propto l^2$) is representative for clumps with column density constant within an order of magnitude and thus seems to be observationally biased \citep{BMc_02}. Also, it holds for clumps that are in equipartition between the gravitational and the kinetic energy or between the gravitational and the magnetic energy \citep{BP_06}. Different considerations of the clump energy budget yield a variety of mass-density relationships, with $-1.5\lesssim x \lesssim 0.4$, as we intend to demonstrate in a forthcoming paper. Moreover, from the perspective of observational clump identification schemes, the Larson-like case $m \propto l^2$ ($x=-0.5$) sets an upper limit on slopes of clump mass-size relationships \citep{Kauf_ea_10a} while analysis of 3D clumps from numerical simulations \citep{Shet_ea_10} increases this limit up to $x\gtrsim0$, depending on the size range. 

The obtained characteristic evolution timescales and SFEs (Table~\ref{tau_sfe}) for the model $x=0.25$ (low PCFE) seem also plausible. In that case, free-fall times of the unstable clumps are about an order of magnitude less than the turbulent crossing times $\tau_{\rm cr}(L)$ (Fig.~\ref{m_tau}). The SFEs are in agreement with estimates of about 5-10 percent in giant MCs \citep{C_ea05}, although the latter could vary in a wider range, depending on the dynamical state of the cloud \citep{CBK08}. Accretion rate calculated according to the chosen description (Sect.~\ref{acc_descr}) also fits the current state of the theory. Numerical simulations of accretion in protostellar clusters by \citet{Bon01a} show that most of the gas is exhausted at $t_{\rm acc}\sim \tau_{\rm ff}$. This sets up only a lower limit of $t_{\rm acc}$ since feedback from massive stars and magnetic fields were neglected in their work but is an appropriate reference value in view of the description of accretion we adopted. In our model, the majority of the unstable clumps at all scales that have given birth to larger clusters ($\gtrsim20$ cores) have consumed the remainining gas at $t=\tau_{\rm ff,\,max}$. Thus the shape of the IMF is practically finished at timescales about $t=\tau_{\rm ff,\,max}$ (Fig.~\ref{imf_025}, top pannels) and further accretion does not influence it significantly.

An apparently realistic IMF is derived from models with $x=0.50$ and $x=0.97$ and higher PCFE (Fig.~\ref{imf_050} and \ref{imf_097}, bottom panels). However, such values of the exponent $x$ cannot be substantiated theoretically and lead to clump mass-size relationships in drastic contradiction both with present observational studies and numerical simulations \citep{Kauf_ea_10b, Shet_ea_10}. Moreover, these models fail to reproduce the substellar IMF -- discontinuity in that mass range is about an order of magnitude larger than found by \citet{TK07} in young Galactic clusters\footnote{If such feature is going to be confirmed.}. 

Accretion on protostellar cores plays a significant role for shaping the final IMF in our model. Therefore it is instructive to make a brief comparison with the recent work of \citet{Dib_ea_10} in which accretion is also one of the key factors and the IMF is also derived from a sum of local core mass distributions. Without going into details, we stress the important conceptual differences. The model of \citet{Dib_ea_10} is aimed at {\it reproduction of the IMF in young clusters} where the SF process has been rapid. The `building blocks' in their approach are the prestellar core mass distributions derived through the formalism of PN02 for different locations within a protocluster clump. Subsequent accretion is also described locally: on cores, injected in the protocluster at a uniform rate and in different epochs, until the gas in the protocluster is dispersed by stellar winds from massive stars. Our model of the IMF is aimed to be {\it representative of star formation in galaxies}, in a wider range of timescales (cf. Fig.~\ref{m_tau}), and its 'building blocks' are unstable clumps generated through a turbulent cascade spanning a range of 3 orders of magnitude in spatial scales. Their fragmentation into prestellar cores proceeds in a self-similar way, depending on the local Jeans mass. The time evolution of the IMF in our model is based on the assumption that the clump population is constantly replenished (Sect.~\ref{treat_timescale}). Unlike the model of \citet{Dib_ea_10}, accretion is spatially averaged for the clump in consideration and feedback from newly formed massive stars is neglected. In view of the large clump statistics assumed in our model, such approach would not affect significantly the predicted IMF although refinements are necessary when the SF process is described specifically at the protocluster scale. 

\section{Summary}
In this work we developed a semi-analytical model of the IMF that takes into account the basic mechanisms in the SF process: gravoturbulent fragmentation and subsequent accretion on the prestellar cores. It is based on the idea of a turbulent cascade that starts from large injection scales $\sim 100~\rm pc$ and transfers energy down to dissipation scales below $0.1~\rm pc$. Turbulence is essentially supersonic at all scales which results in network of interacting shocks and formation of condensations (clumps) with lognormal distribution of densities. The main assumption in our model is the existence of a power-law relation between the clump mass and the clump density $n\propto m^x$, where $x$ is a scale-free parameter. It leads to lognormal clump mass distribution and to a scale-dependent critical mass that determines the fraction of gravitationally unstable clumps (if any) at given scale. Summation of their mass distributions over all scales yields their composite CMF: asymmetric with shallow power-law part (slope $-0.8\lesssim\Gamma\lesssim-0.6$), for $x<-0.5$, and steeper, with a Salpeter-like slope ($\Gamma\sim -1.2$), for $x\ge-0.5$. The fragmentation of unstable clumps is assumed to produce a lognormal mass distribution of prestellar cores, centred around the clump Jeans mass. Adopting constant core formation efficiency $\epsilon$, we let the rest of the clump mass $(1-\epsilon)m$ to be distributed on the formed cores through competitive accretion. The duration of this process is within $\sim1$ clump free-fall times $\tau_{\rm ff}(x)$ as the gas density in the protocluster approaches an $r^{-2}$ profile. Eventually, the IMF is derived through summation over all clumps with halted accretion, build at all scales, and allowing constant replenishment of clumps with faster evolution.

The free parameters of the proposed IMF model are the exponent $x$ in the clump mass - clump density relation, the turbulent forcing parameter $b$ and the core formation efficiency $\epsilon$. Models with $x=0.25$ and low $\epsilon$ turn out to be consistent with the observational IMF. The physics behind the proposed relation $n\sim m^x$ is to be substantiated. Fragmentation of unstable clumps is another element of the model that needs further elaboration. As demonstrated by \citet{Kl01}, it depends on the turbulent injection scale and hence may yield a variety of mass distributions of dense cores. Eventually, we point out that the implemented description of competitive accretion within a fragmented clump gives only general estimates. Accurate modelling of this complex process is practically impossible in an analytical or semi-analytical framework. 

To conclude, the proposed model reproduces correctly the effects of gravoturbulent fragmentation and competitive accretion on the IMF at galactic scales. It is based on reasonable physical assumptions about the SF process and should be considered as a first step toward a more extensive theoretical framework.

\section*{Acknowledgements}
We thank our referee, Paolo Padoan, whose critical approach and comments helped us to improve this work.\\
T.V. acknowledges support by the {\em Deutsche Forschungsgemeinschaft} (DFG) under grant KL 1358/9-1 and by the Scientific Research Foundation, Ministry of Education and Sciences, Bulgaria, under contract VU-F-201/06. Part of the calculations were performed at the cluster Physon at the Faculty of Physics in Sofia, Bulgaria, under contract VU-205/06.\\
R.S.K.\ acknowledges financial support from the {\em Landesstiftung Baden-W{\"u}rttemberg} via their program International Collaboration II (grant P-LS-SPII/18) and from the German {\em Bundesministerium f\"{u}r Bildung und Forschung} via the ASTRONET project STAR FORMAT (grant 05A09VHA). R.S.K. furthermore gives thanks for subsidies from the DFG under grants no.\ KL 1358/1, KL 1358/4, KL 1358/5, KL 1358/10, and KL 1358/11, as well as from a Frontier grant of Heidelberg University sponsored by the German Excellence Initiative.

\label{lastpage}

\end{document}